\documentclass{ckm}

\def\BR{\text{BR}}

\usepackage{amsmath,amssymb}
\usepackage{txfonts}

\confname{Workshop on the CKM Unitarity Triangle, IPPP Durham, April 2003}

\title{Rare Decays of D Mesons}

\author{Svjetlana Fajfer\addressmark{a}\addressmark{b}%
\thanks{e-mail: \texttt{svjetlana.fajfer@ijs.si}}}
\address[a]{Department for theoretical physics, Jo\v{z}ef Stefan Institute,
Slovenia}
\address[b]{Department for physics, Faculty for mathematics and physics,
University of Ljubljana, Slovenia}

\begin{document}

\begin{abstract}%
We analyze the dominant decay mechanisms in the radiative decays $\rm D\to V
\gamma$, $\rm D\to P(V)\ell^+\ell^-$, $\rm D\to\gamma\gamma$, $\rm
D\to\ell^+\ell^-\gamma$, and we discuss possible chances to observe the
effects of new physics in these decays. Then we comment on Cabibbo
allowed $\rm D\to K\pi\gamma$ decays with nonresonant $\rm K\pi$ and
we probe the role of light vector mesons in these decays.
\end{abstract}

\maketitle

The K and B meson rare decays offer a chance to investigate effects of
new physics in the down-like quark sector, while the rare D decays
offer an opportunity to investigate the FCNC effects in the up-like
sector. For the rare D decays only upper bounds have been established
so far for a sizable number of the radiative or dilepton weak decay of
D mesons \cite{CLEO1,CLEO2,FOCUS,E791}. New chances to observe these
decay rates will appear soon at BELLE, BABAR and Tevatron. In analysis
\cite{FPS2,FPS3,FPS1,FSZ,FSZ1} of the dominant long distance
contributions in $\rm D\to V\gamma$, $\rm D\to P(V)\ell^+\ell^-$, $\rm
D\to\gamma\gamma$, and $\rm D\to\ell^+\ell^-\gamma$ were determined.
Possibilities to observe the effects of new physics were discussed.

In the standard model (SM) the $\rm c\to u\gamma$ transition is
strongly GIM suppressed at one loop electroweak order giving a 
branching ratio of order $10^{-18}$ \cite{BGHP}. The QCD corrected
effective Lagrangian gives $\rm\BR(c\to u\gamma)\simeq3\times10^{-8}$
\cite{GMW}. A variety of models beyond the standard model were
investigated and it was found that the gluino exchange diagrams
\cite{PW} within general minimal supersymmetric SM (MSSM) give the
largest enhancement
\begin{equation}
\rm\frac{\BR(c\to u\gamma)_{\text{MSSM}}}{\BR(c\to u\gamma)_{\text{SM}}} 
\simeq10^2.
\label{1}
\end{equation}
The $\rm c\to u\ell^+\ell^-$ amplitude is given by the $\gamma$ and Z
penguin diagrams and W box diagram at one-loop electroweak order in
the SM, and is dominated by the light quark contributions in the
loop. We have used renormalization group improved effective weak
Lagrangian within SM \cite{FSZ1} and we found that the rates are
$\rm\BR(c\to ue^+e^-)_{\text{SM}}=2.4\times10^{-10}$ and $\rm\BR(c\to
u\mu^+\mu^-)_{\text{SM}}=0.5\times10^{-10}$. However, the present
bounds on MSSM parameters lead to $\rm\BR(c\to
ue^+e^-)_{\text{MSSM}}=6.0\times10^{-8}$ and
$\rm\BR(c\to u\mu^+\mu^-)_{\text{MSSM}}=2.0\times10^{-8}$ \cite{FPS1}.

Long distance physics overshadows such effects since it usually
dominates the decay amplitude \cite{FPS2}. The long distance
contribution is induced by the effective nonleptonic $|\Delta c|=1$
weak Lagrangian
\begin{align}
{\cal L}=&-A_{ij}\big[a_1\bar u\gamma^{\mu}(1-\gamma_5)q_i
\bar q_j\gamma_{\mu}(1-\gamma_5)c
\nonumber\\
&+a_2\bar q_j\gamma_{\mu}(1-\gamma_5)q_i\bar
u\gamma^{\mu}(1-\gamma_5)c\big],
\label{eff}
\end{align}
with $A_{ij}=(G_F/\sqrt{2})\;V_{cq_j}^*V_{uq_i}$, accompanied by the
emission of the virtual photon. Here $q_{i,j}$ denote the d or s quark
fields. The effective Wilson coefficients are $a_1=1.2$ and $a_2=-0.5$
\cite{FPS2}. In our calculations of the long distance effects we use
\cite{FPS2,FPS3,FPS1} the theoretical framework of heavy meson chiral
Lagrangian \cite{wise}. The factorization approximation has been used
for the calculation of weak transition elements. In Table \ref{tab1}
we give the branching ratios of $\rm D\to V\gamma$ decays \cite{FPS2}.
The uncertainty is due to relative unknown phases of various
contributions.
\begin{table}[h]
\begin{center}
\caption{The branching ratios for $\rm D\to V\gamma$ decays.}
\label{tab1}
\begin{tabular}{ll}
\hline
$\rm D\to V\gamma$ & $\BR$ \\
\hline
$\rm D^0 \to{\bar K}^{*0}\gamma$ &$[6-36]\times10^{-5}$ \\
$\rm D_s^+\to\rho^+\gamma$ &$[20-80]\times10^{-5}$ \\
$\rm D^0\to\rho^{0}\gamma$&$[0.1-1]\times10^{-5}$ \\
$\rm D^0\to\omega\gamma$ &$[0.1-0.9]\times10^{-5}$  \\  
$\rm D^0 \to \Phi \gamma$ &$ [0.4 - 1.9 ]\times 10^{-5} $ \\
$\rm D^+ \to \rho^+ \gamma$ &$ [0.4 -6.3]\times 10^{-5}$\\
$\rm D_s^+ \to K^{*+ }\gamma$ &$[1.2 - 5.1]\times 10^{-5}$ \\
$\rm D^+ \to K^{*+} \gamma$ &$ [0.3- 4.4]\times 10^{-6}$ \\
$\rm D^0 \to K^{*0} \gamma$ &$ [0.3 - 2.0] \times 10^{-6}$   \\ 
\hline
\end{tabular}
\end{center}
\end{table}
Although the branching ratios are dominated by the long distance
contributions, the size of the short distance contribution can be
extracted from the difference of the decay widths
$\rm\Gamma(D^0\to\rho^{0}\gamma)$ and $\rm\Gamma(D^0\to\omega\gamma)$
\cite{FPSW}. Namely, the long distance mechanism $\rm c\bar u\to d\bar
d\gamma$ screens the $\rm c\bar u\to u\bar u\gamma$ transition in $\rm
D^0\to\rho^{0}\gamma$ and $\rm D^0\to\omega\gamma$, the $\rho^{0}$ and
$\omega$ mesons being mixture of $\rm u\bar u$ and $\rm d\bar d$.
Fortunately, the LD contributions are mostly canceled in the ratio
\begin{align}
R&=\frac{\BR({\rm D}^0\to\rho^{0}\gamma)-\BR({\rm D}^0\to\omega\gamma)}
{\BR({\rm D}^0\to\omega\gamma)}
\nonumber\\
&\propto\operatorname{Re}\frac{A({\rm D^0\to u\bar u\gamma})}
{A({\rm D^0\to d\bar d\gamma})},
\label{2}
\end{align}
which is proportional to the SD amplitude $A({\rm D^0\to u\bar
u\gamma})$ driven by $\rm c\to u\gamma$. This ratio
$R_{\text{SM}}=6\pm15\%$ in the Ref.\ \cite{FPSW}, and can be enhanced
up to ${\cal O}(1)$ in the MSSM. In addition to the $\rm c\to u\gamma$
searches in the charm meson decays, we have suggested to search for
this transition in $\rm B_c\to B_u^*\gamma$ decay \cite{FPS-B}, where
the long distance contribution is much smaller.

The amplitudes for exclusive decays $\rm D\to V\ell^+\ell^-$ and $\rm
D\to P\ell^+\ell^-$ are dominated by the long distance contributions.
The branching ratios for these decays, as obtained in
\cite{FPS3,FPS1}, are given in Tables \ref{tab3} and \ref{tab4}. In
these Tables the first column represents the SD contributions, while
in the second column the rates coming from the LD contributions are
given. The rates for $\rm D\to Ve^+e^-$ are comparable to those in
Table \ref{tab3} and can be found in \cite{FPS3}.
\begin{table}[h]
\begin{center}
\caption{The $\rm D\to V\mu^+\mu^-$ branching ratios.}
\label{tab3}
\begin{tabular}{lll}
\hline
$\rm D\to V\mu^+\mu^-$& $\BR_{\text{SD}}$ & $\BR_{\text{LD}}$\\
\hline
 $\rm D^0\to \bar K^{*0}\mu^+ \mu^- $ & $0$& 
$[1.6 - 1.9]\times 10^{-6}$\\
 $\rm D^+_s\to \rho^+\mu^+ \mu^- $ & $0$& 
$[3.0- 3.3]\times 10^{-5}$\\
 $\rm D^0\to \rho^{0}\mu^+ \mu^- $ & $9.7 \times 10^{-10}$& 
$[3.5-4.7]\times 10^{-7}$\\
 $\rm D^0\to \omega\mu^+ \mu^- $ & $9.1 \times 10^{-10}$& 
$[3.3-4.5]\times 10^{-7}$\\
 $\rm D^0\to \phi \mu^+ \mu^- $ & $0$& 
$[6.5-9.0] \times 10^{-8}$\\
 $\rm D^+\to \rho^+ \mu^+ \mu^- $ & $4.8 \times 10^{-9}$& 
[$1.5-1.8] \times 10^{-6}$\\
 $\rm D^+_s\to K^{*+} \mu^+ \mu^- $ & $1.6 \times 10^{-9}$& 
$ [5.0 -7.0] \times 10^{-7}$\\
 $\rm D^+\to K^{*+} \mu^+ \mu^- $ & $ 0$& 
$ [3.1-3.7]\times 10^{-8}$\\
$\rm D^0\to \ K^{*0}\mu^+ \mu^- $ & $0$& 
$[4.4-5.1] \times 10^{-9}$\\
\hline
\end{tabular}
\end{center}
\end{table}

\begin{table}[h]
\begin{center}
\caption{The $\rm D\to P\ell^+\ell^-$, ($\rm\ell=e$, $\mu$) branching ratios.}
\label{tab4}
\begin{tabular}{lll}
\hline
$\rm D\to P \ell^+\ell^-$& $\BR_{\text{SD}}$ & $\BR_{\text{LD}}$\\
\hline
 $\rm D^0\to \bar K^{0}\ell^+\ell^-$ & $0$& 
$4.3\times 10^{-7}$\\
 $\rm D^+_s\to \pi^+\ell^+ \ell^-$ & $0$& 
$ 6.1\times 10^{-6}$\\
 $\rm D^0\to \pi^{0}\ell^+ \ell^- $ & $1.9 \times 10^{-9}$& 
$2.1 \times 10^{-7}$\\
 $\rm D^0\to \eta \ell^+ \ell^-$ & $2.5 \times 10^{-10}$& 
$4.9\times 10^{-8}$\\
 $\rm D^0\to \eta' \ell^+ \ell^-$ & $ 9.7 \times 10^{-12} $& 
$2.4 \times 10^{-8}$\\
 $\rm D^+\to \pi^+ \ell^+ \ell^-$ & $9.4\times 10^{-9}$& 
$1.0 \times 10^{-6}$\\
 $\rm D^+_s\to K^{+} \ell^+ \ell^-$ & $ 9.0\times 10^{-10}$& 
$ 4.3 \times 10^{-8}$\\
$\rm D^+\to K^{+} \ell^+ \ell^-$ & $ 0$& 
$ 7.1 \times 10^{-9}$\\
$\rm D^0\to \ K^{0}\ell^+ \ell^- $ & $0$& 
$1.1 \times 10^{-9}$\\
\hline
\end{tabular}
\end{center}
\end{table}

The experimental upper bounds for $\rm D \to P \ell^+\ell^-$ decay
rates were recently improved by FOCUS \cite{FOCUS}, with new upper
bounds of $10^{-5}$ or less, close to the theoretical predictions
\cite{FPS1,BGHP0}. The allowed kinematic region for the dilepton mass
$m_{\ell\ell}$ in the $\rm D\to P\ell^+\ell^-$ decay is
$m_{\ell\ell}=[2m_\ell,m_{\rm D}-m_{\rm P}]$. The long distance
contribution has resonant shape with poles at
$m_{\ell\ell}=m_{\rho^0}$, $m_\omega$, $m_\phi$ \cite{FPS1,BGHP0}.
There is no pole at $m_{\ell\ell}=0$ since the decay $\rm D\to
P\gamma$ is forbidden. The short distance contribution is rather flat.
The spectra of $\rm D\to P e^+e^-$ and $\rm D\to P\mu^+\mu^-$ decays
in terms of $m_{\ell\ell}$ are practically identical. The difference
in their rates due to the kinematic region $m_{\ell\ell}=[2m_{\rm e},2m_\mu]$ is
small and we do not consider them separately.

The differential distribution for $\rm D^{+,0} \to \pi^{+0} \ell^+ \ell^-$
\cite{FPS1,BGHP0} indicates that the high mass dilepton region might
yield an opportunity for detecting $\rm c \to u \ell^+ \ell^-$.  The
kinematics of the processes $\rm D \to V \ell^+\ell^-$ would be more
favorable to probe the possible supersymmetric enhancement at small
$m_{\ell\ell}$, but the long distance contributions in these channels
are even more prominent \cite{FPS3}.

We undertook an investigation of the $\rm D^0\to \gamma \gamma$ decay
\cite{FSZ} and recently CLEO gave the first upper limit $\rm\BR(D^0\to
\gamma \gamma)< 2.9 \times 10^{-5}$ \cite{CLEOG}. The short distance
contribution is expected to be rather small and therefore the main
contribution is coming from long distance interactions. The total
amplitude is dominated by terms proportional to $a_1$ that contribute
only through loops with Goldstone bosons. At this order the amplitude
receives an annihilation type contribution proportional to the $a_2$
Wilson coefficient, given by the Wess-Zumino anomalous term coupling
light pseudoscalars to two photons. Terms which contain the anomalous
electromagnetic coupling of the heavy quark Lagrangian are suppressed
compared to the leading loop effects \cite{FSZ}. The invariant
amplitude for $\rm D^0 \to \gamma\gamma $ decay can be decomposed
using gauge and Lorentz invariance into a parity conserving and parity violating parts
\cite{FSZ}. Within this framework \cite{FSZ}, the leading
contributions are found to arise from the charged $\pi$ and K mesons
running in the chiral loops. As a result of these internal exchanges
our calculation predicts that the $\rm D\to 2 \gamma$ decay is mostly
a parity violating transition \cite{FSZ}. We find that the predicted
branching ratio is
\begin{equation}
\BR({\rm D}^0\to \gamma \gamma)= (1.0 \pm 0.5)\times 10^{-8}. 
\label{fin-res}
\end{equation}
The short distance contribution within the SM has been estimated in
\cite{BGHP0}, which found that it cannot exceed the long distance part.

Recently we have studied the dominant contributions to the rare decay
mode $\rm D \to \ell^+ \ell^- \gamma$ \cite{FSZ1}. The long distance
contributions can be divided into nonresonant and resonant
contributions. The first is the result of $\pi$, K loops, while the
second one occurs on account of the $\rm D^0 \to V \gamma \to
\ell^+ \ell^-\gamma$ transition where $\rm V = \rho^0$, $\omega$,
$\pi$ vector mesons. We have calculated that nonresonant
contributions give the rates
\begin{eqnarray}
&&\BR({\rm D^0\to \gamma e^+ e^-})_{\text{nonres}} = 1.29 \times 10^{-10},
\nonumber\\
&&\BR({\rm D^0\to \gamma \mu^+ \mu^-})_{\text{nonres}} = 0.21 \times 10^{-10}.
 \label{llgnr}
\end{eqnarray}
The resonant contributions are 
\begin{eqnarray}
&&\BR({\rm D^0\to \rho \gamma \to \ell^+ \ell^- \gamma})\simeq 5 \times 10^{-11},
\nonumber\\
&&\BR({\rm D^0\to \omega \gamma \to \ell^+ \ell^- \gamma})\simeq 8 \times 10^{-11},
\nonumber\\
&&\BR({\rm D^0\to \phi \gamma \to \ell^+ \ell^- \gamma})\simeq 1\times 10^{-9}.
\label{llgr}
\end{eqnarray}
As seen from these results the largest contributions come from the
$\phi$ resonance. Finally, combining both contributions we estimate
that the dominant long distance contributions in these decays give
\begin{eqnarray}
&&\BR({\rm D^0\to \gamma e^+ e^-})_{\text{SM}} = 1.2\times 10^{-10},
\nonumber\\
&&\BR({\rm D^0\to \gamma \mu^+ \mu^-})_{\text{SM}} = 1.1 \times 10^{-10}.
 \label{llgsm}
\end{eqnarray}
Using simplest supersymmetric extension of SM we investigate possible
effects. The largest contribution is expected from gluino-squark
exchanges \cite{FSZ1,PW}. Using single mass insertion we find
\cite{FSZ1},
\begin{eqnarray}
&&\BR({\rm D^0\to \gamma e^+ e^-})_{\text{SM}} = 1.4\times 10^{-9},
\nonumber\\
&&\BR({\rm D^0\to \gamma \mu^+ \mu^-})_{\text{SM}} = 1.2 \times 10^{-9}.
 \label{llgmssm}
\end{eqnarray}
The situation rather changes if $\cal R$-parity violation is allowed
as in \cite{FSZ1}. The effective interaction is then
\begin{equation}
{\cal L}_{\text{eff}} = \frac{ \tilde \lambda'_{i2k} 
 \tilde \lambda'_{i1k}}{2 M_{\tilde {\rm D}{\cal R}}^2} 
 (\bar {\rm u}_L \gamma^{\mu} c_L)(\bar \ell_L \gamma^{\mu} \ell_L),
\label{Rparity}
\end{equation}
where $\tilde \lambda'_{ijk}$ are the coefficients of the $\cal
R$-parity breaking terms of the superpotential \cite{FSZ1}. Using new
experimental bound $\BR({\rm D^+ \to \pi^+ \mu^+ \mu^-}) < 8.8\times
10^{-6}$ \cite{FOCUS} we obtain a new bound
\begin{equation}
\tilde \lambda^{\prime}_{i2k} 
 \tilde \lambda^{\prime}_{i1k}\leq 0.003
 \left[\frac{ M_{\tilde {\rm D} {\cal R}}^2}{100\,{\rm GeV}}\right]^2.
\label{Rbound}
\end{equation}
Using this bound and allowing photon emission from $\ell^+$ and
$\ell^-$, with photon energy cut $E_\gamma \geq 50\,{\rm MeV}$ and
$E_\gamma\geq 100\,{\rm MeV}$
\begin{eqnarray}
&&\BR({\rm D^0\to \gamma e^+ e^-})_{E_\gamma \geq 50\,{\rm MeV}}^{\cal R} = 
4.5\times 10^{-9},
\nonumber\\
&&\BR({\rm D^0\to \gamma \mu^+ \mu^-})_{E_\gamma \geq 50\,{\rm
MeV}}^{\cal R} = 
50 \times 10^{-9},
\nonumber\\
&&\BR({\rm D^0\to \gamma e^+ e^-})_{E_\gamma \geq 100\,{\rm
MeV}}^{\cal R} = 
4.5\times 10^{-9},
\nonumber\\
&&\BR({\rm D^0\to \gamma \mu^+ \mu^-})_{E_\gamma \geq 50\,{\rm
MeV}}^{\cal R} = 
46 \times 10^{-9}.
 \label{llgR}
\end{eqnarray}
Note that the SM predictions are not affected by the cuts on the soft
photon energy at the order $E_\gamma \geq 100\,{\rm MeV}$ as the bulk
of the contribution either comes from the resonances or the low $p^2$
region \cite{FSZ1}. The cut on $E_\gamma$ is the cut on the high $p^2$
region.

In addition to the D meson rare decays in which the FCNC transitions
might occur, we have studied the Cabibbo allowed radiative decays $\rm
D^+\to{\bar K}^0\pi^+\gamma$, $\rm D^0 \to K^- \pi^+\gamma$ and $\rm D^0
\to {\bar K}^0 P$ with $\rm P = \pi^0$, $\eta$, $\eta'$ with nonresonant K
$\pi$, which we consider to be the most likely candidates for early
detection. Here we used the heavy quark chiral Lagrangian supplemented
by light vector mesons, as the theoretical framework.

The radiative $\rm K \to \pi \pi \gamma$ decays offer 
useful information in the study of the K meson decays. Their counterparts
in charm sector are charm meson decays into $\rm P_1P_2\gamma$. Using
the factorization approximation for the calculation of weak transition
elements we use the information obtained from semileptonic decays
\cite{castwo,BFO2}. The nonleptonic $\rm D \to K\pi$ amplitude cannot be
calculated accurately in the factorization approximation from the
diagrams provided by our model. Such a calculation gives a rather good
result for the $\rm D^+ \to {\bar K}^0\pi^+$ channel but is less
successful for the $\rm D^0 \to K^+ \pi^-$ decay. In order to overcome
this deficiency and to be able to present accurately the
bremsstrahlung component of the radiative transition, we shall use an
alternative approach. We take the
experimental values for the $\rm D \to K \pi$ amplitudes, in the
calculation of the bremsstrahlung component. In order to accommodate
this, we write \cite{FAS} the decay amplitude as
\begin{align}
{\cal M} = &-\frac{G_f}{\sqrt{2}}V_{cs}V_{du}^*
\{F_0 [\frac{q \cdot\varepsilon}{q  \cdot k}-
\frac{p\cdot \varepsilon}{p\cdot k}]
\nonumber\\
&+F_1 [ (q \cdot\varepsilon) (p  \cdot k)- 
(p \cdot\varepsilon) (q  \cdot k)]
\nonumber\\
&+F_2 \varepsilon^{\mu \alpha \beta \gamma}
\varepsilon_\mu v_\alpha k_\beta q_\gamma \},
\label{amp1}
\end{align}
where $F_0$ is the experimentally determined $\rm D \to K\pi$
amplitude and $F_1$, $F_2$ are the form factors of the electric and
magnetic direct transitions which we calculate with our model. When
intermediate states appear to be on the mass shell, we use
Breit-Wigner formula.  Thus, we get \cite{FAS} for the branching
ratios of the electric transitions including bremsstrahlung, with
$|F_0|$ determined experimentally and taking the photon energy cut
$E_c\geq100\,{\rm MeV}$
\begin{equation}
\BR({\rm D^+\to\bar K^0\pi^+\gamma})_{\text{PV,ex}}^{E_c}
=(2.3-2.5)\times10^{-4}.
\label{rpvp2}
\end{equation}
For the $\rm D^0$ radiative decay we get
\begin{equation}
\BR({\rm D^0 \to K^- \pi^+ \gamma})_{\text{PV,ex}}^{E_c}
=(4.3-6.0)\times10^{-4}.
\label{rpv02}
\end{equation}
The uncertainty in the $F_0/F_1$ phase is less of a problem in $\rm D^+
\to \bar K^0 \pi^+ \gamma$ than in $\rm D^0 \to K^- \pi^+ \gamma$. If we
take the bremsstrahlung amplitude alone as determined from the
knowledge of $|F_0|$, disregarding the direct electric $F_1$ term, the
above numbers are replaced by $2.3\times 10^{-4}$ for $\rm D^+$ decay
and $5.5\times 10^{-4}$ for the $\rm D^0$ decay. The contribution of
the direct parity violating term (putting $F_0 = 0$), is $\BR({\rm
D^+\to \bar K^0 \pi^+ \gamma})_{\text{dir,PV} }=1.0 \times 10^{-5}$
and $\BR({\rm D^0 \to K^-\pi^+ \gamma})_{\text{dir,PV}}= 1.64 \times
10^{-4}$. For the parity conserving direct magnetic transition we get
\cite{FAS}
\begin{align}
&\BR({\rm D^+ \to \bar K^0 \pi^+ \gamma})_{\text{PC}} = 
2.0 \times 10^{-5}, 
\label{rpvp2m}
\\
&\BR({\rm D^0 \to K^- \pi^+ \gamma})_{\text{PC}} = 
1.4\times 10^{-4}.
\label{rpv02m}
\end{align}
Hence, the two direct transitions are predicted to be of about the
same strength. If we disregard the contribution of vector mesons to
the direct part of the radiative decays, the parity-conserving part of
the amplitude is considerably decreased, by one order of magnitude in
the rate in $\rm D^+ \to \bar K^0 \pi^+ \gamma$ decay and by two
orders of magnitude in $\rm D^0 \to K^- \pi^+ \gamma$. On the other
hand, their contribution is not felt in a significant way in the
parity -- violating part of the amplitudes. In any case, the detection
of the direct part of these decays at the predicted rates, will
constitute a proof of the important role of the light vector mesons.

The decay mode $\rm D^0$ with neutral pseudoscalar mesons has no
bremsstrahlung part of the amplitude. The light virtual vector mesons
give the main contribution to the decay amplitude.  The decay $\rm D^0
\to {\bar K^0} \pi \gamma$ is predicted \cite{FAS} to have the rate
\begin{equation}
\BR({\rm D^0 \to {\bar K^0} \pi \gamma}) =3.4 \times 10^{-4},
\label{D-n}
\end{equation}
with comparable contributions from the parity-conserving and
parity-violating parts. The decays $\rm D^0 \to {\bar K^0} \eta (\eta')
\gamma$ are expected with rates $1.1 \times10^{-5}$ and $0.4
\times10^{-7}$, respectively, and are mainly parity conserving.

In summary: we find that the best candidate to search for new physics
effects in charm decay seems to be the difference $\BR({\rm
D^0\to\rho^{0} \gamma}) - \BR({\rm D^0 \to \omega \gamma})$. The
observation of the rate of $\BR({\rm D}^0\to \gamma \gamma)$ larger
than $10^{-8}$ can signal new physics, while the rate $\BR({\rm
D}^0\to\mu^+\mu^-\gamma)$ of the order $10^{-8}$ can be explained as a
result of the $\cal R$-parity violating MSSM effects. We hope that
interesting features of these decays result in an experimental search
in the near future.

\end{document}